\documentclass[epsf]{mn2e}
\usepackage{graphicx}
\def\mmm{(m-M)$_0$}
\def\ebv{E($B-V$)~}

\def\gsim{\;\lower.6ex\hbox{$\sim$}\kern-7.75pt\raise.65ex\hbox{$>$}\;}
\def\lsim{\;\lower.6ex\hbox{$\sim$}\kern-7.75pt\raise.65ex\hbox{$<$}\;}

\title[Berkeley 22]{Berkeley 22, an old and distant open cluster towards 
 the Galactic anticentre} 

\author[Di Fabrizio et al.]{L. Di Fabrizio$^1$, A. Bragaglia$^2$,
        M. Tosi$^2$, G. Marconi$^{3,4}$ \\
 $^1$ INAF - Telescopio Nazionale Galileo,
 38700 Santa Cruz de La Palma, Spain,
 e-mail difabrizio@tng.iac.es,	\\
 $^2$ INAF - Osservatorio Astronomico di Bologna - Via Ranzani 1, I-40127 
      Bologna, Italy,
      e-mail angela.bragaglia, monica.tosi @bo.astro.it \\
  $^3$ INAF - Osservatorio Astronomico di Roma - Via dell'Osservatorio 5, 
      I-00040 Monte Porzio, Italy, \\
  $^4$ ESO, Alonso de Cordova 3107, Vitacura, Santiago, Chile,
      e-mail gmarconi@eso.org}

\date{}

\begin{document}
\maketitle

\begin{abstract}
We present deep CCD $BVI$ photometry of the distant, old open cluster Berkeley 22,
covering from the red giants branch (RGB) to about 6 magnitudes below the main
sequence (MS) turn-off.
Using the synthetic Colour - Magnitude Diagram method with three different
types of stellar evolutionary tracks, we estimate values and theoretical
uncertainty of distance modulus \mmm, reddening \ebv, age $\tau$ and 
approximate metallicity. The best fit to the data is obtained for  
13.8 $\leq$ \mmm $\leq$ 14.1, 0.64 $\leq$ \ebv $\leq$ 0.65, 
2.0 $\leq$ $\tau$ $\leq$ 2.5 Gyr (depending on the amount of overshooting from 
convective regions adopted in the stellar models) and solar metallicity.

\end{abstract}


\section{Introduction}

This paper is part of our general program of observations of galactic open
clusters (the Bologna Open Clusters Chemical Evolution project, or
BOCCE, Bragaglia \& Tosi 2004, and in preparation). These aggregates, and in
particular the old ones,  represent an ideal tool to study the properties of
the disc of our Galaxy and their evolution with time (e.g. Janes 1979; 
Panagia \& Tosi 1981; Friel 1995). 
Open clusters with ages less than about 5 Gyr are also
good tests of the stellar evolutionary tracks, since they can be used to
discriminate  between models with or without overshooting, and on its supposed
extension.

In our well tested approach, we obtain the cluster colour-magnitude diagram
(CMD) from deep and precise photometry, and derive at
the same time age, distance, reddening and approximate metallicity using the
synthetic CMD method (Tosi et al. 1991). We
have already published our findings for 15 clusters (see Kalirai \& Tosi 
2004; Tosi et al. 2004, and references therein).
Whenever possible,  we also try to derive precise metal abundances with high
resolution spectroscopy (Bragaglia et al. 2001; Carretta et al. 2004). 

We present here the photometric results on the relatively unstudied open cluster 
Berkeley 22 (C0555+078), with coordinates $\alpha_{2000} = 5^h58^m27^s$,
$\delta_{2000}= +07^o 45\arcmin 28\arcsec$, l = $199^o.80$, b = $-8^o.05$. 

\begin{figure}
\begin{center}
\includegraphics[bb=95 188 513 603, clip, scale=0.55]{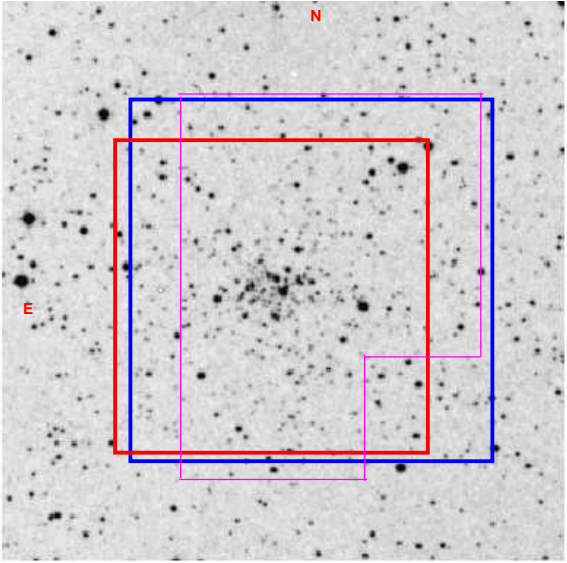}
\caption{Map of our field: the image, taken from the DSS, is 10$\times$10
arcmin$^2$, and is oriented with North up and East left. The two square boxes
(thick lines) represent the Danish (the larger field) and the SuSI2 pointings;
the approximate position of Kaluzny's composite field is indicated by the
thinner line polygon. The external field, not shown here, is 
about 30 arcmin south of this position.}
\label{fig-map}
\end{center}
\end{figure}

The best colour magnitude diagram published so far for this object is by
Kaluzny (1994, hereafter K94). He observed a small region around the cluster
centre (covering with a mosaic of 5 pointings less than 4.5 arcmin
by 6 arcmin, see Fig.~\ref{fig-map}) with the KPNO 2.1m telescope, in
the $B$, $V$, and $I$ filters. He obtained quite deep CMDs  and derived a
very small apparent radius for the cluster, of about 2 arcmin on the sky.
Kaluzny estimated age, reddening, and distance (age $\simeq$ 3 Gyr, \ebv =
0.65, and distance from the Sun = 6 kpc), assuming solar metallicity. This
assumption was made for lack of any precise information on the true metal
abundance, but he argued that the actual metallicity should be lower 
than solar, because of the morphology of the red giant clump 
(in which case, reddening and distance estimates would increase).

Phelps, Janes \& Montgomery (1994, their fig. 13) also observed this
cluster, in the $V$ and $I$ filters. From their data, of quality inferior to
K94, they estimate an age of 2.1 Gyr, based on the MAI (Morphological Age
Indicator, i.e. the difference in magnitude between the MS
turn-off and the red clump). In a companion paper Janes \& Phelps (1994)
give for this cluster E(B-V)=0.65, as deduced from the mean colour of the
red clump.

Salaris, Weiss \& Percival (2004), using a new calibration of the
morphological age index,  and adopting $\delta$V = 2.1 from 
Friel (1995) and [Fe/H] = $-0.30$ from
Gratton (2000), find for Be 22 an age of 4.26 $\pm$ 1.65 Gyr.

Lata et al. (2002) give integrated magnitude and colour for Be 22,
based on published data: following their age - magnitude relation, an age
of 1.1 Gyr is derived, while the use of the age - colour relation produces
an unreasonable value (28 Gyr).

The maps of dust infrared emission by Schlegel, Finkbeiner \& Davis (1998)
indicate a reddening of \ebv = 0.65, while Dutra \& Bica (2000) 
find \ebv = 0.63, in very good agreement.
Noriega-Mendoza and Ruelas-Mayorga (1997), applying a technique similar to
the Sarajedini (1994) method for simultaneous metallicity and reddening
determination, find  [Fe/H] = $-0.42$ and \ebv = 0.70.
The reddening seems in close agreement with the other literature values,
even if tied to a metallicity different from K94. 

Finally, Be 22 has recently been suggested to be part of the Canis Major
cluster family (Frinchaboy et al. 2004, Martin et al. 2004), as many other
OCs. 

In Section 2 we describe our data and the reduction process, in Section
3 we present our CMDs and compare them to literature ones, 
in Section 4 we derive the
cluster parameters, and a summary is given in Section 5.

\begin{figure}
\begin{center}
\includegraphics[bb=45 185 290 575, scale=0.8]{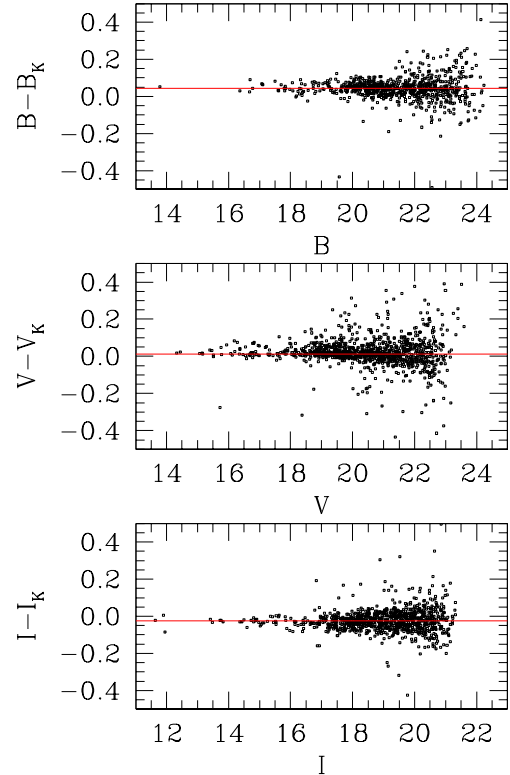}
\caption{Comparison between our photometry and K94: the mean differences (our
magnitudes minus K94) are: +0.045 in B, +0.011 in V, and $-$0.026 in I.}
\label{fig-confrok}
\end{center}
\end{figure}

\section{Observations and data reduction}

Berkeley 22 was observed at three different telescopes. Table 1 gives a 
log of the three runs and Fig.~\ref{fig-map} shows the used observed fields (as
well as K94's).

The first data were obtained with the 1.54m Danish telescope located in La
Silla, Chile, on 1995  March 6, 8, 10. We used the direct camera with CCD
No 28 (Tek 1024$\times$1024, scale 0.377 arcsec pix$^{-1}$, field of view 6.4
$\times$ 6.4 arcmin$^2$). We took only one field, centred on the cluster (see
Fig.~\ref{fig-map}). Exposures ranged from 300 to 1500 seconds in B, from 20 to
900 seconds in V and  from 30 to 900 seconds in I. 
Conditions were photometric, and seeing ranged between  0.9 and 1.4
arcsecarcsec.
Standard stars areas were observed, and later used to calibrate the whole
set of data on the cluster.

We later reobserved the cluster on 2000 October 2 and 3 with the Telescopio
Nazionale Galileo (TNG) on Roque de los Muchachos, mounting DOLORES (Device
Optimized for LOw RESolution) a focal reducer instrument equipped with a Loral
thinned, back-illuminated 2048$\times$2048 CCD, with scale 0.275 arcsec
pix$^{-1}$, and field of view of 9.4 $\times$ 9.4  arcmin$^2$. Seeing
conditions were not ideal for this rather crowded cluster, and we had some
technical problems, which plagued chiefly the long exposures on the central
field. We decided to  retain for further analysis only the data obtained for
the external comparison field, located 30 arcmin South of the cluster centre,
and calibrated independently using standard areas observed during the run.

Finally, deep exposures (B: 3 $\times$ 600 sec; V: 2 $\times$ 300 sec,  I: 2
$\times$ 300 sec) on the central field were obtained with the Superb Seeing
Imager 2 (SuSI2) on the New Technology Telescope (NTT), on 2000 December 1 (see
Fig.~\ref{fig-map}). The two EEV CCDs cover a field of view of  5.4 $\times$
5.4 arcmin$^2$, and a gap of about 8 arcsec is present between the two chips;
we binned the chips 2$\times$2,  yielding a scale of 0.16 arcsec pix$^{-1}$.
Seeing was better than about  1 arcsec.

\begin{table*}
\begin{center}
\caption{Log of observations; Ra and Dec are in units of hours, minutes,
seconds, and degrees, primes, seconds respectively; exposure times are in
seconds.}
\begin{tabular}{lccc}
\hline
Date   & Instrument  & Ra, Dec & Exposures \\
\hline
1995 March 6,8,10 &Danish + direct CCD & 05:58:10, +07:45:17  & B=300,1500; V=20,60,300,720,900; I=30,60,180,480,900\\
2000 October 3    & TNG + DOLORES      & 05:58:27, +07:15:28  & B=60,900;   V=10,300; I=10,300\\
2000 December 1   &NTT + SuSI2         & 05:58:28, +07:45:32  & B=600,600,600; V=300,300; I=300,300\\
\hline
\end{tabular}
\end{center}
\end{table*}

For all sets of data, standard CCD reduction of bias subtraction, trimming, and
flatfield correction were performed. We applied to all frames the usual
procedure  for point spread function (PSF) study and fitting available in
DAOPHOT--II (Stetson 1987, Davis 1994) in IRAF\footnote{
IRAF is distributed by the National Optical Astronomical Observatory, which 
are operated by the Association of Universities for Research in Astronomy, under 
contract with the National Science Foundation} 
environment.
All frames were searched independently (the two SuSi2 chips were reduced 
separately), and the output catalogues were later aligned in coordinates.
Magnitudes for the frames centred on Be 22 were  aligned  to a reference frame
for each band, chosen among the Danish ones, since they were acquired under
photometric conditions. A colour term is present in the B filter and was taken
into account in the photometry calibration.  All the catalogues of instrumental
magnitudes (5 in $B$, 7 in $V$ and $I$) were correlated;  multiple measures of
the same star in the same filter were averaged using a $2\sigma$ rejection
cycle.  Finally, aperture photometry was performed on isolated stars  on the  
reference images (47 stars for the B, 51 for the V and 40 for the I bands
respectively), to compute a correction to the PSF derived magnitudes and be on
the same system as the photometric standard stars. This correction is -0.240,
-0.223, -0.141 mag in $B$, $V$ and $I$ respectively.  Analysis of the external
field proceeded in a completely analogous way.

The calibration from instrumental magnitudes to the Johnson-Cousins standard
system was obtained on the Danish data in the case of the central field, using 
the same set of equations derived for Pismis 2 (Di Fabrizio et al. 2001), 
a cluster observed in the same run. We refer to that paper for all details.
The TNG data for the external field were calibrated separately, using the
following equations, derived from the Stetson (2000) standard areas 
Markarian A and PG0231+051:

$$ B = b +0.0580 \cdot (b-v) -3.5576 ~~~~(r.m.s.= 0.014) $$
$$ V = v -0.1707 \cdot (b-v) -3.7182 ~~~~(r.m.s.= 0.019) $$
$$ V = v -0.0963 \cdot (v-i) -3.7481 ~~~~(r.m.s.= 0.016) $$
$$ I = i +0.0394 \cdot (v-i) -4.1932 ~~~~(r.m.s.= 0.024) $$

where $b, v, i$ are the instrumental magnitudes and $B, V, I$ are the 
corresponding Johnson-Cousins magnitudes.

We tested the completeness of our photometry in the B, V and I bands on the
deepest frames (the SuSI2 oned for Be 22), using the procedure  developed by P.
Montegriffo at the Bologna Observatory. Artificial stars were added, about 120
at a time,  to the frames at random positions properly chosen in order not to
alter the  crowding conditions. The images were then reduced exactly as before,
finding objects and measuring magnitudes using the same parameters, PSF and
selection criteria. The process was repeated as many times as to reach about
50000 added artificial stars, to ensure statistical significance. We then
derived the completeness level (see Table \ref{tabcompl}) as the ratio of
recovered over added stars in different magnitude bins. A a star is considered
as recovered if it is found within 0.5 pixel and 0.75 mag of the simulated one.
The former value is due to the pointing precision and the latter corresponds to
the magnitude difference resulting from complete overlap of two equal brightness
stars. By keeping all the artificial stars with lower magnitude difference, we
can trace the trend of blending with magnitude (see e.g. Tosi et al. 2001) and
introduce it in the synthetic CMD procedure described below. 

The differences between the input and output magnitudes of the artificial stars
provide a good estimate of the photometric errors (and possible image blends).
The errors in the cluster field turn out to be smaller than 0.01 mag for 
$B\leq$18.2, $V$ and $I \leq$ 16.5, and larger that 0.1 mag for $B\geq$24.2, 
$V\geq$23.5 and $I \geq$ 21.5.

\begin{table}
\begin{center}
\caption{Completenes in the three bands}
\begin{tabular}{cccc}
\hline
mag      &Compl$_B$  &Compl$_V$ &Compl$_I$    \\ 
\hline  
15.250 & 1.000 & 1.000 & 0.996 \\ 
15.750 & 1.000 & 1.000 & 0.996 \\ 
16.250 & 0.989 & 0.998 & 0.995 \\ 
16.750 & 0.989 & 0.997 & 0.995 \\ 
17.250 & 0.989 & 0.999 & 0.996 \\ 
17.750 & 0.989 & 0.997 & 0.989 \\ 
18.250 & 0.987 & 0.997 & 0.994 \\ 
18.750 & 0.985 & 0.995 & 0.986 \\ 
19.250 & 0.983 & 0.991 & 0.984 \\ 
19.750 & 0.979 & 0.987 & 0.982 \\ 
20.250 & 0.976 & 0.986 & 0.977 \\ 
20.750 & 0.972 & 0.981 & 0.965 \\ 
21.250 & 0.967 & 0.975 & 0.940 \\ 
21.750 & 0.957 & 0.969 & 0.800 \\ 
22.250 & 0.935 & 0.950 & 0.425 \\ 
22.750 & 0.892 & 0.882 & 0.085 \\ 
23.250 & 0.818 & 0.780 & 0.008 \\ 
23.750 & 0.689 & 0.505 & 0.001 \\ 
24.250 & 0.488 & 0.111 & 0.000 \\ 
24.750 & 0.264 & 0.009 & 0.000 \\ 
25.250 & 0.103 & 0.001 & 0.000 \\ 
25.750 & 0.029 & 0.000 & 0.000 \\ 
26.250 & 0.007 & 0.000 & 0.000 \\ 
\hline
\end{tabular}
\end{center}
\label{tabcompl}
\end{table}

\section{The colour - magnitude diagram}

\begin{figure*}
\begin{center}
\includegraphics[bb= 25 230 590 590,scale=0.8]{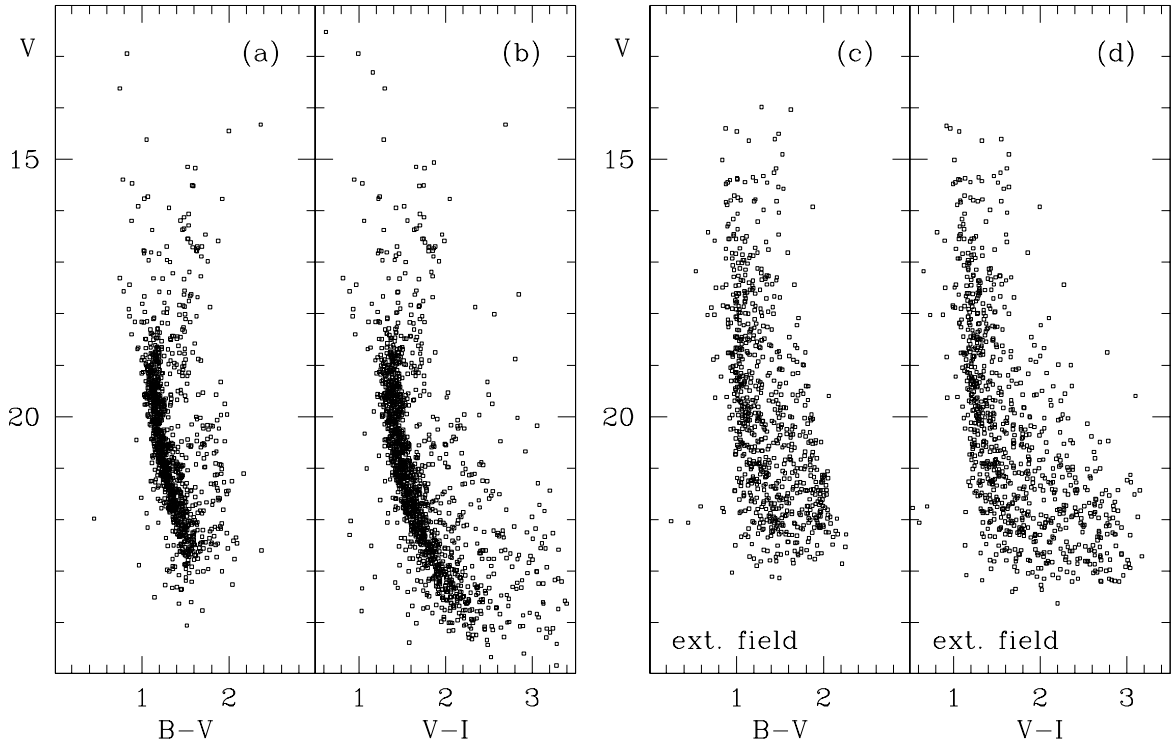}
\caption{(a) and (b) CMDs for Be 22, on a field of 6.4 $\times$ 6.4 arcmin$^2$;
(c) and (d) CMDs for the external field, covering  9.4 $\times$ 9.4
arcmin$^2$.}
\label{fig-cmd}
\end{center}
\end{figure*}

\begin{figure*}
\begin{center}
\includegraphics[bb=70 275 530 680, scale=0.8]{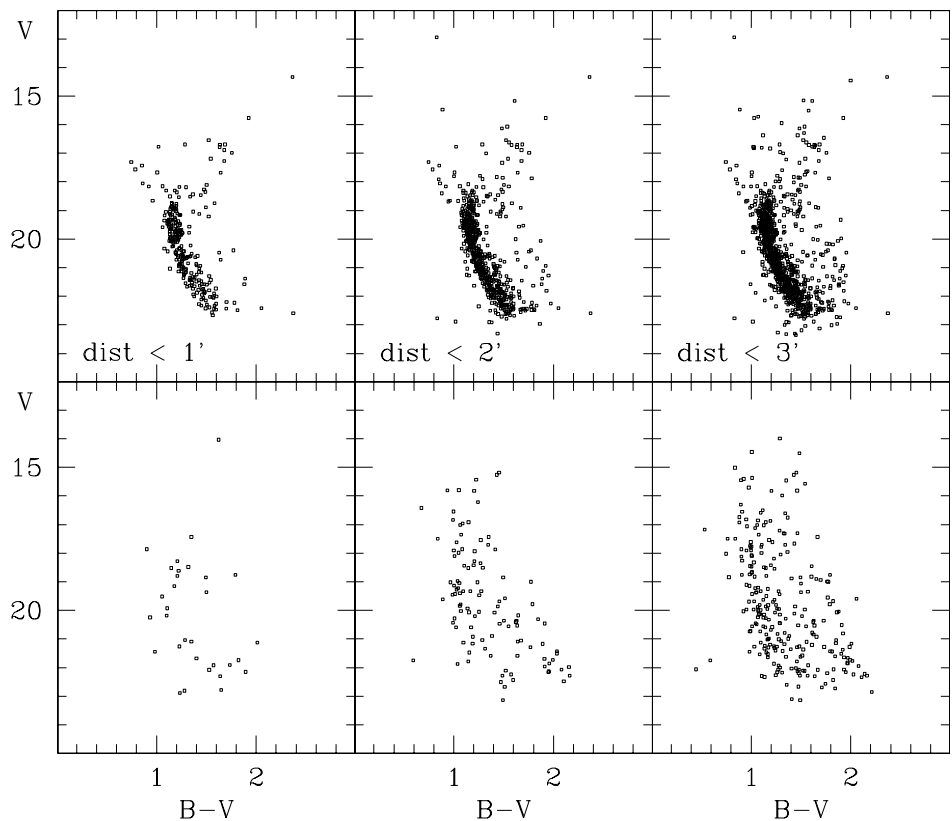}
\caption{Upper panels: radial CMDs for Be22, showing all star within a radius
of 1 arcmin, 2 arcmin, 3 arcmin. Lower panels: stars in corresponding areas of
the external field.}
\label{fig-rad}
\end{center}
\end{figure*}

\begin{figure*}
\includegraphics[bb=70 450 530 680, scale=0.8]{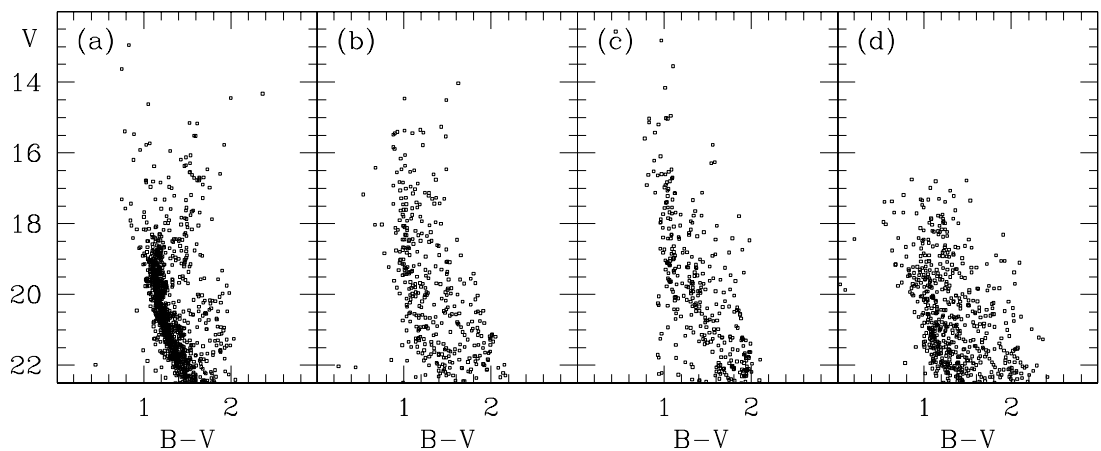}
\begin{center}
\caption{(a) Our central field; (b) a corresponding area of the
external comparison field; (c) results of the Besan\c con
model on the same area, and with the same incompleteness factors as in 
our external field; (d) the Canis Major system (Bellazzini et al. 2004),
also taken on our same area, and 
shifted in accordance to the different reddenings.}
\label{fig-model}
\end{center}
\end{figure*}

The final cluster catalogue contains 1494 stars, of which 1123 have $B$, $V$,
and $I$ magnitudes, 4 have only $B$ and $V$ magnitudes, and 367 have only $V$
and $I$ magnitudes.
The catalogue was astrometrized using private software written by P.
Montegriffo at the Bologna Observatory; pixel coordinates were transformed to
equatorial coordinates by means of 140 objects in common with the Guide Star
Catalogue 2;  residuals of the transformation between the two systems are of
about 0.08 arcsec both in right ascension and declination.
Tables containing the photometry, the pixel and equatorial coordinates will
be available in electronic form  through the BDA\footnote{
http://obswww.unige.ch/webda/webda.html} (Base Des Amas, Mermilliod 1995).

We also identified stars in common between our sample and K94, and a comparison
of the two photometries is shown in Fig. \ref{fig-confrok}: they appear to be 
in reasonable agreement. There is no trend with magnitude, and the
average zero point shifts (our photometry minus K94) are of +0.045 mag in $B$,
+0.011 mag in $V$, and $-0.026$ mag in $I$.  

The CMDs for Be 22 based on our photometry are shown in Fig. 
\ref{fig-cmd}(a,b), while Fig. \ref{fig-cmd}(c,d) show
the external comparison field (note that the 2 CMDs are based on a 6.4 $\times$
6.4 arcmin$^2$ and 9.4 $\times$
9.4 arcmin$^2$ fields, respectively).
The cluster sequences are very well delineated, with a main sequence 
extending from $V\simeq  19$ and $B-V \simeq 1.05$ or $V-I \simeq 1.25$ 
at the turn-off (TO) point 
down to $V \simeq$ 23 or 24.5, for about 5 - 6 magnitudes. The subgiant
and red giant branches are also visible, even if suffering more
from  field contamination.
The red clump, i.e., the locus of core-He burning stars, is present at
$V \simeq 16.65$  and $B-V \simeq 1.55$, $V-I \simeq 1.80$.

The field contamination shows up both as a component redder than the cluster MS,
which is easily identifiable, and a second one, perfectly overimposed to
the cluster (see Fig. \ref{fig-cmd}). 
To try to separate Be 22 from the background we have produced radial
CMDs (Fig. \ref{fig-rad}, with centre at Ra = $5^h58^m27^s$,
Dec = $+07^o 45\arcmin 28\arcsec$), showing 
stars within a given radius, and a portion
of the comparison field of same area.
The similarity between the resulting MS's is striking, given 
that the comparison field is 30 arcmin away from the cluster centre.

A further check of the field component has been done comparing our
fields with the Galaxy model by Robin et al. (2003, available
at the web site {\tt www.obs-besancon.fr/www/modele}).
We have retrieved the model for an area equivalent to our central field of view
located at l  =199.8, b = $-8.1$. 
In Fig. ~\ref{fig-model}(a,b,c) we show Be 22, the
external field,  and the results of the Besan\c con model (to which we have
applied our incompleteness factors)  respectively, all for the same area. We
notice immediately an eccess of stars in our external field with respect to the
model; they follow the same distribution of Be 22 MS.

Since Be 22 has been proposed, although with some doubts, to be
related to the Canis Major galaxy (Martin et al. 2004, Frinchaboy et al. 2004),
we have compared our data also to these recently presented by Bellazzini
et al. (2004, kindly made available in electronic form by M. Bellazzini),
shown in Fig. \ref{fig-model}(d), after correcting for the different 
reddenings, and for an area equal to ours. 
Again, there are similarities, but the Canis Major sequences
are bluer than our cluster.

\begin{figure}
\begin{center}
\includegraphics[bb=35 180 550 680,scale=0.45]{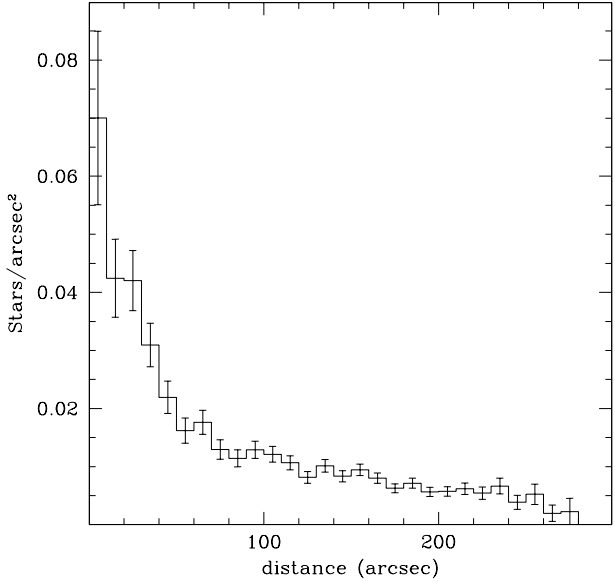}
\caption{Projected density distribution of Be 22 stars.}
\label{fig-annuli}
\end{center}
\end{figure}

Judging from Fig. \ref{fig-map} the cluster appears to have a small radius 
($\sim$ a few arcmin, see also K94), but  its radial extent is not easily
determined from our data. We have computed the radial star density distribution
(in annuli  10 arcsec wide), and the profile is shown in Fig.
~\ref{fig-annuli}.
We do not see any strong evidence of
flattening, so we can not determine the point where the cluster disappears. 
This is also consistent with the presence of cluster stars in  the
external field.

In literature, Be 22 has \ebv $\simeq$ 0.65, without anyone ever suggesting 
it to be differential. This is what we find also 
from our data (see next Section): the MS width is perfectly compatible with
a single value of reddening, once photometric errors and binaries are
taken into account.

\begin{figure*}
\begin{center}
\includegraphics[bb=80 180 545 715]{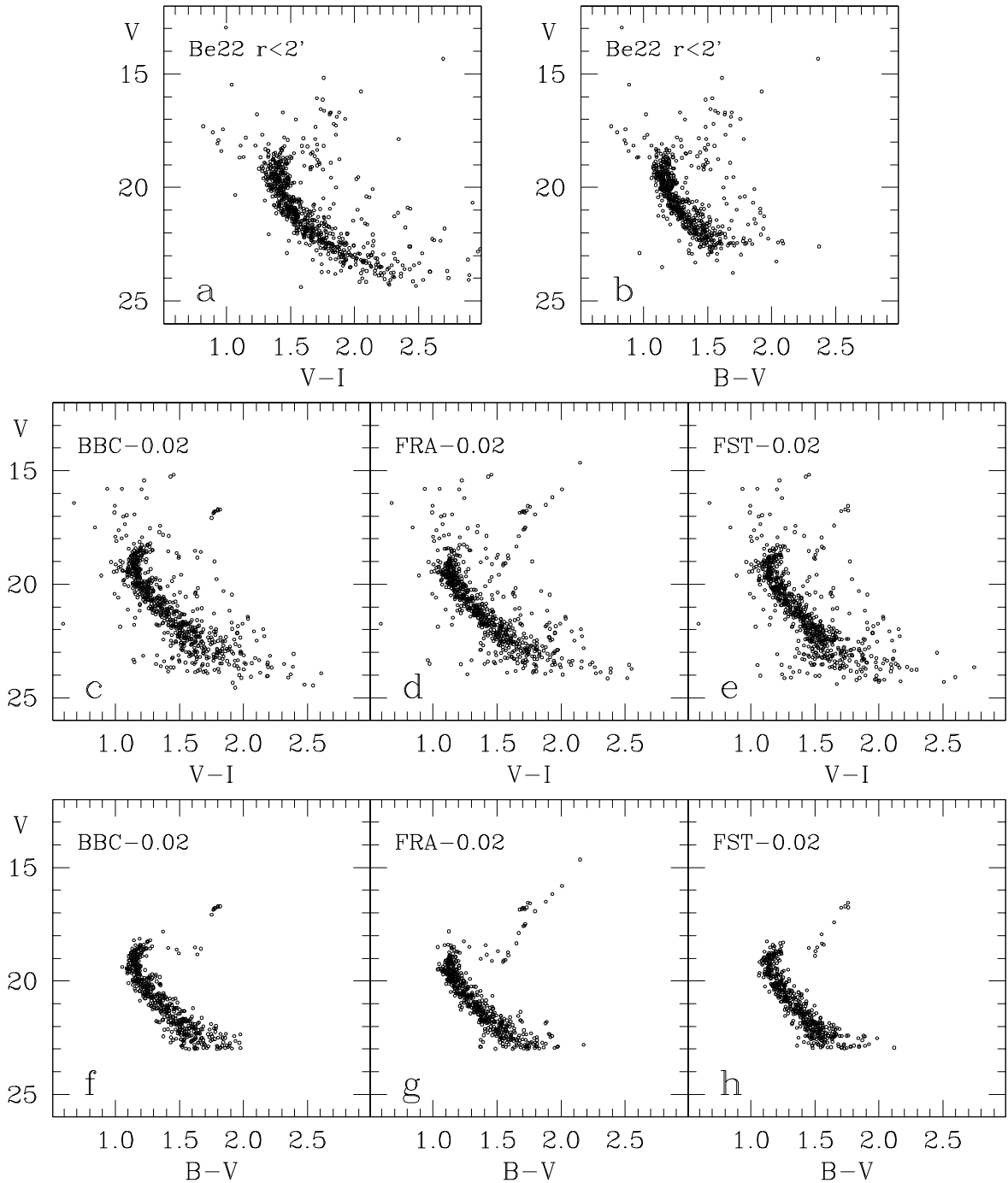}
\caption{Top panels: empirical CMDs of the region of Be 22 within 2 arcmin
from the cluster centre. 
Panels c through h: corresponding
synthetic CMDs in better agreement with the data for the three
different types of stellar models (labelled on the top left corner of each
box together with the model metallicity). The BBC model of panels c and f 
assumes age = 2.4 Gyr, \ebv=0.64, \mmm=13.8. The FRA model of panels d and g 
assumes age = 2.1 Gyr, \ebv=0.65, \mmm=14.0. The FST model of panels e and h 
assumes age = 2.5 Gyr, \ebv=0.65, \mmm=13.9.} 
\label{sim}
\end{center}
\end{figure*}

\section{Cluster parameters}

For sake of homogeneity, age, distance and reddening of all the clusters of 
the BOCCE project are derived with the same synthetic CMD method (Tosi et al.
1991) and the same sets of stellar evolution models (see  
Kalirai \& Tosi 2004 and references therein). The best values of the parameters  
are found by selecting the cases providing synthetic CMDs with morphology, 
colours, number of stars in the various evolutionary phases and luminosity     
functions (LFs) in better agreement with the observational ones.
To test the effect of different input physics on the derived  
parameters, we run the simulations with three different types of stellar   
evolutionary tracks, assuming different prescriptions for the treatment of 
convection and ranging from no overshooting to  high overshooting from 
convective regions. 

Be 22 has very uncertain metallicity estimates (see the Introduction), 
hence, we have created the synthetic CMDs adopting, for each type of 
stellar models, various metallicities. For sake of homogeneity, we keep
considering as solar metallicity tracks those with Z=0.02 (i.e. the ones 
calibrated on the Sun by their authors), although the actual solar metallicity 
is supposed to be lower (see Asplund et al. 2004).

The adopted sets of stellar tracks are listed in Table 3, where the 
corresponding references are also given, as well as the  model metallicity and
the information on their overshooting assumptions. 
The transformations from the theoretical luminosity and effective temperature 
to the Johnson-Cousins magnitudes and colours have been performed using Bessell, 
Castelli \& Pletz (1998) conversion tables and assuming E$(V-I)$ = 1.25 \ebv
(Dean et al. 1978) for all sets of models. Hence, the different results
obtained with different stellar models must be ascribed fully to the models
themselves and not to the photometric conversions.
Different conversion tables can lead to different magnitudes of the synthetic
stars and affect our results on the cluster reddening and metallicity. We have
however checked that the effect on colours of adopting other conversions is
smaller than, or at most comparable to, the photometric errors and not
sufficient to introduce metallicity mismatches.

\begin{table}
\begin{center}
\caption{Stellar evolution models adopted for the synthetic CMDs;
the
FST models actually adopted here are an updated version of the published ones
(Ventura, private communication)}
\vspace{5mm}
\begin{tabular}{cccl}
\hline\hline
Set & metallicity & overshooting & Reference \\
\hline
BBC & 0.02  & yes         & Bressan et al. 1993   \\
BBC & 0.008 & yes         & Fagotto et al. 1994   \\
FRA & 0.02  & no          & Dominguez et al. 1999 \\
FRA & 0.01  & no          & Dominguez et al. 1999 \\
FRA & 0.006 & no          & Dominguez et al. 1999 \\
FST & 0.02  & $\eta$=0.02 & Ventura et al. 1998   \\
FST & 0.02  & $\eta$=0.03 & Ventura et al. 1998   \\
FST & 0.006 & $\eta$=0.02 & Ventura et al. 1998   \\
FST & 0.006 & $\eta$=0.03 & Ventura et al. 1998   \\
\hline
\end{tabular}
\end{center}
\label{models}
\end{table}

The synthetic stars are attributed the photometric error derived from 
the artificial stars' tests performed on the actual images. They are
retained in (or excluded from) the synthetic CMD according to the  
completeness  factors listed in Table \ref{tabcompl}.
All the synthetic CMDs have been computed either assuming that all the cluster
stars are single objects or that a fraction of them are members of binary 
systems with random mass ratio. 
We find, as in many other clusters, that a 
binary fraction around 30\% well reproduces the observed distribution along the
main sequence. All the synthetic CMDs shown in the figures assume this 
fraction of binaries. These models reproduce well also the observed colour
distribution, thus suggesting that differential reddening does not affect 
the examined regions of the cluster.

\begin{figure*}
\begin{center}
\includegraphics[bb=95 180 560 426]{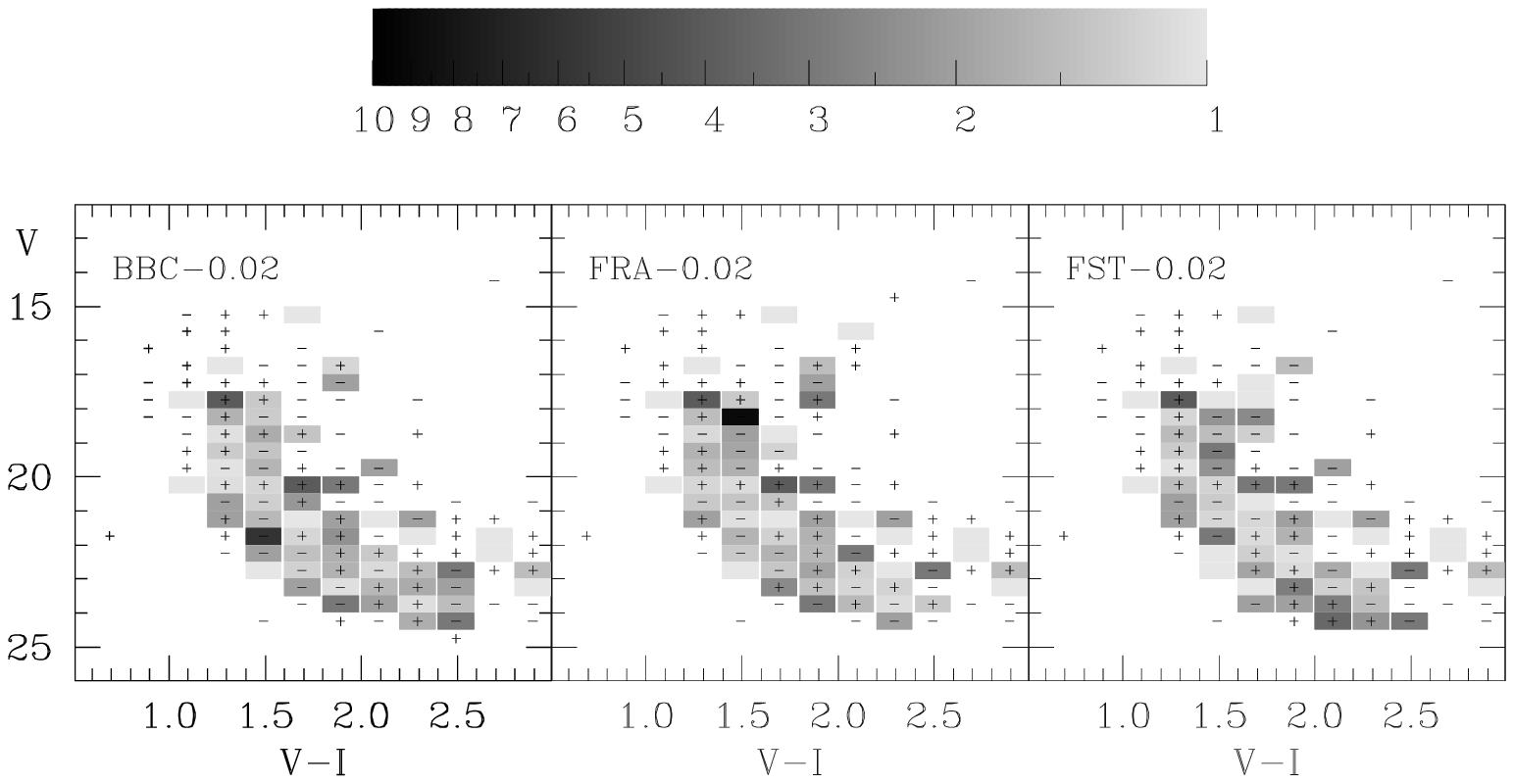}
\caption{Comparison of the theoretical and empirical CMDs of Fig.~\ref{sim}:
ratio of synthetic to observed star counts.
In each cell of the CMDs the grey-scale level indicates  the ratio 
$s/o$ when a plus is shown or $o/s$ when a minus is shown. Cells with no
sign have $s/o$=1. Cells with no grey filling and with a plus contain 
synthetic stars but no observed ones. Vice versa, cells with no grey filling and
with a minus contain observed stars but no synthetic ones. 
The grey-scale coding is shown at the top. See text for further details.} 
\label{simobsratio}
\end{center}
\end{figure*}

To our knowledge, no studies have been performed on the cluster membership.
To reduce the field contamination, without limiting excessively the number
of measured stars, we have used as reference for the simulations the $V, V-I$ 
CMD of the cluster region within a radius of 2 arcmin from the centre. 
This central region contains 744 stars with measured $V$ and $I$ magnitudes. 
The corresponding CMD is shown in Fig.~\ref{sim}(a). 
Since an equivalent area of the external field contains 131 stars, we can 
assume that the cluster members in the central region are presumably 613. 
The synthetic CMDs  have therefore  been created with this number 
of objects.
To compare the resulting CMDs and LFs with the empirical ones, we have added to
the 613 theoretical stars the 131 objects measured in the external field, which
wouldn't be reproduceable by the models.

Due to the very red colour of Be22, and in spite of the depth of the B frames, 
only 600 of the 744 central stars
have also a measured $B$ magnitude (see  Fig.~\ref{sim} b). We used 
the synthetic $V, B-V$ CMDs to check the colours and the morphology 
of the various evolutionary sequences.

We find that with all the adopted sets of stellar tracks only the solar 
metallicity models are able to reproduce both in $B-V$ and in $V-I$ the 
colours of the observed CMD sequences. All the other metallicities lead 
to extended SGBs and excessively red RGBs (once fitting the 
MS colours), or wrong slopes of the MS, or non self-consistent $B-V$ and 
$V-I$ MS colours. 

The synthetic CMDs in better agreement with the data for each of the three types
of models are shown in Fig.~\ref{sim}, where the $V, V-I$ diagrams of panels c,
d and e show the 613 synthetic stars plus the 131 ones of the external field. 
In the $V, B-V$ CMDs of panels f, g and h only the synthetic stars with
magnitude compatible with the measured B magnitudes are plotted.  
All the synthetic CMDs seem to underpredict the colour extension of
the clump, but the lack of information on the actual cluster membership of
the objects observed in that CMD region prevents any firm conclusion on this
point.

To appreciate the quality of the agreement between the synthetic and the
observational diagrams of Fig.~\ref{sim}, we have divided the CMDs in cells and
compared the number of synthetic ($s$) and observed ($o$) stars counted in 
each cell.
Fig.~\ref{simobsratio} displays the ratio $s/o$, if $s/o \geq 1$, or $o/s$,
if $s/o <1$ with a logarithmic grey-scale coding. The cell size is 0.5 mag 
in $V$ and 0.2 mag in $V, V-I$. The grey level is lightest for $s/o$=1 and
becomes increasingly dark for increasing difference in either direction of
synthetic and observed counts. To keep memory of the actual $s/o$, a plus
sign is indicated in all cells where  $s > o$ and a minus sign in those where 
$s < o$.  Cells with no grey filling  have no observed stars when
a plus is plotted, and have no synthetic stars when a minus is plotted. 

\begin{figure}
\begin{center}
\includegraphics[bb=120 230 445 715, scale=0.7]{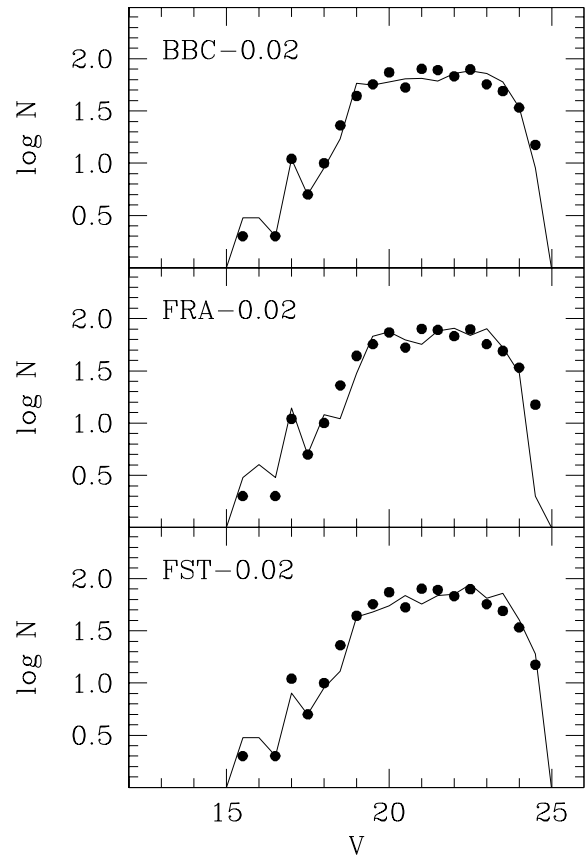}
\caption{Luminosity functions of the selected stars in the cluster region 
within 2 arcmin from the center (dots). The lines correspond to the best
synthetic models for each type of tracks, whose CMDs are shown in Fig.
\ref{sim}.} 
\label{simlf}
\end{center}
\end{figure}

The LFs of the best cases for the BBC, FRA and FST models are shown in 
Fig.~\ref{simlf} (lines) and compared to that of the CMD of Fig.~\ref{sim}(a).

The BBC solar models reproduce quite well the shape and colours of the observed
evolutionary sequences, but only for $\tau$=2.4 Gyr [Fig.~\ref{sim}(c) and (f)]
they predict not only the right luminosity but also the right number of 
clump and RGB stars. The LF (top panel in Fig.~\ref{simlf}) is in excellent
agreement with the data as well.  At lower metallicity the BBC models 
provide more extended SGBs (hence, redder RGBs) and turn-off  hooks, that
make their synthetic CMDs less satisfactory than the solar metallicity ones. 

The FST solar models also reproduce well the data, if the intermediate 
overshooting case $\eta=0.02$ is chosen. The luminosity and the number of 
stars of the clump are consistent with the data for whatever age between 
2.3 and 2.7 Gyr, but only for $\tau$=2.5 Gyr is the TO shape in excellent 
agreement with the observational one [Fig.~\ref{sim}(e) and (h)]. 
As apparent from the LF of the bottom panel of Fig.~\ref{simlf}, the relative
number of stars in the different evolutionary phases is slightly less correct
than that of the BBC best case, but still very good. On the other hand, these
tracks reproduce the observed RGB colour (both in $B-V$ and in $V-I$ better than
the others.
If the higher overshooting, $\eta=0.03$, models are adopted, the clump is always
less populated and the TO tends to come out excessively hooked, independently of
the age. At lower metallicity, the FST models predict  MS slopes deviating from
the observed one, specially in the $V, B-V$ CMD. 

The FRA solar models allow for the best clump reproduction, but have pronounced
TO hook and tend to populate the whole RGB, contrary to what is observed. The
best compromise is obtained with $\tau$=2.0 or 2.1 Gyr and, respectively,
\mmm=14.1 or 14.0, with the same \ebv=0.65 [Fig.~\ref{sim}(d) and (g)]. At lower
metallicity the TO shape worsens, the MS curvature diminishes and the RGB is
overpopulated and too red. 

The results obtained with the various sets of models are quite consistent with
each other, once the known differences in the input physics are taken into
account. In fact, we obtain the youngest best fitting age $\tau$ (2.0 or 2.1 
Gyr) with the FRA models, without overshooting, the oldest one (2.7 Gyr) with 
the FST$-\eta=0.03$ ones with maximum overshooting, and intermediate values 
with the sets with intermediate overshooting (2.4 Gyr with the BBC models 
and 2.5 with the FST$-\eta=0.02$ models). The reddening \ebv is derived with a 
very small uncertainty (only $\pm$ 0.005), and the distance modulus \mmm~ with 
an uncertainty of $\pm$ 0.15. If we consider only the two best fitting models
(BBC and FST$-\eta=0.02$), the agreement is even more striking, with $\tau$=2.4
and  2.5 Gyr, \ebv=0.64 and 0.65, and \mmm=13.8 and 13.9, respectively.

\section{Summary and discussion}

The application of the synthetic CMD method to Be 22 has allowed us to derive
very narrow ranges for the values of its age, distance and reddening. The
theoretical uncertainty related to the adoption of different stellar models is
significant only on the age, where it inevitably reflects the different
overshooting
assumptions. Even allowing for a less selective choice of the best cases, the
reddening \ebv always turns out between 0.64 and 0.65, in excellent agreement
with the derivations by K94, Janes \& Phelps (1994), and the maps by Schlegel et
al. (1998). Higher reddenings, of course, are inferred if a metallicity lower
than solar is adopted. The BBC models with Z=0.008, the FRA models with Z=0.01
and the FST ones with Z=0.006 all need \ebv=0.70, the value suggested by
Noriega-Mendoza \& Ruelas-Mayorga (1997), but none of them reproduces
the observed features as well as the solar tracks. We thus favour a solar
metallicity for this cluster, despite the existing suggestions for lower
abundances. 

In fact this cluster could be an important piece in the puzzle regarding the
abundance distribution with Galactocentric distance: while almost anyone thinks
in term of a radial gradient, Twarog, Ashman \& Anthony-Twarog (1997) and more
recently Corder \& Twarog (2001) presented the case for a step distribution,
with clusters closer than 10 kpc from the centre  having  metallicities
different from those of more external objects.  Be 22 is farther than 10 kpc,
and from our study appears to have solar abundance; if that is confirmed, it
would represent a flaw in Twarog et al.'s (1997) reasoning.  However, none of
the presently available metallicity derivations resides on firm grounds, and we
believe that high-resolution spectroscopy is needed for a safe estimate. 

As discussed in the previous Section, the distance modulus  formally ranges
between 13.8 and 14.1, but the very best models indicate \mmm=$13.85 \pm 0.05$,
corresponding to a distance from the Sun of $5.89 \pm 0.14$ kpc and from the
Galactic centre of $13.66 \pm 0.14$ kpc. This places Be 22 among the outermost
old open clusters, second in our BOCCE sample only to Be 29, which is however
much farther out, at $22.0 \pm 0.6$ kpc (Tosi et al. 2004,  Carraro et
al. 2004) from the centre. 

Although affected by a substantial reddening, Be 22 does not show evidence for
differential effects. The CMDs of different cluster regions do not present 
colour shifts, and the synthetic diagrams have the same colour spread of the 
empirical ones by simply adopting the photometric errors.

Evaporation of low mass stars does not appear to have significantly affected
Be 22, contrary to what happens to many open clusters. The synthetic 
luminosity functions match very well the observational ones both in the inner
regions and in the whole surveyed field. Since all the stars born in the
cluster, and still alive at the present epoch, which passed the incompleteness 
tests, are counted in the synthetic LF, this good match indicates that, 
within the uncertainties, only a minor fraction of the objects born in the 
examined area can have moved away during the cluster lifetime. Yet, the
comparison of the CMD of our external field with those of the cluster field and
of the Besan\c con galactic models suggests that MS stars like those of Be22 and
not expected from the models are found also at 30 arcmin from the cluster 
centre, corresponding, at our derived distance to 52 pc. 
Is this just a coincidence, or has Be 22 really been
able to form stars over such a large area ?

Be 22 has been suggested to belong to the family of the debris of the newly 
discovered Canis Major satellite (Frinchaboy et al. 2004, Martin et al. 2004),
and we have made a direct comparison to it (Fig. ~\ref{fig-model}), finding
Be 22 redder than CMa, once considered the different absorptions.
Our fields of view are probably too small to address this issue, but the
circumstance that neither the CMD of the cluster nor that of our external 
field appear to contain alien contaminating structures, when compared with 
the predictions of the Besan\c con Galactic model, suggests that Be 22 
is out of significant streams or overdensity regions. 
Interestingly enough, we found the same lack of evidence for Be 29,
which was also suggested by Frinchaboy et al. to be connected to Canis Major.
Admittedly, lack of evidence is not evidence of lack.

\bigskip\bigskip\noindent 
ACKNOWLEDGEMENTS

This paper is based on observations collected at the European Southern 
Observatory, Chile, and at the Italian Telescopio Nazionale Galileo.
Most of L.D.F.'s work was done while at the Osservatorio Astronomico di 
Bologna.
We warmly thank P. Montegriffo, whose programs were used in the data analysis,
M. Bellazzini for interesting discussions and for providing the 
Canis Major data,
P. Ventura, who kindly provided the unpublished FST models, and L. Angeretti
for help with the comparison procedures. 
We appreciated the constructive comments of the anonymous referee.
The bulk of the 
numerical code for CMD simulations was originally written by L. Greggio. 
We acknowledge the use of the valuable BDA database, maintained by J.C.
Mermilliod, Geneva.
This research has made use of the Simbad database, operated at CDS,
Strasbourg, France.

\end{document}